\documentclass[aps,twocolumn,amsmath,amssymb,pre,floatfix]{revtex4-1}
\usepackage{amssymb}
\usepackage{amsmath}
\DeclareMathOperator{\sech}{sech}
\usepackage{graphicx}
\usepackage{epsfig}
\usepackage[latin1]{inputenc}
\usepackage{dcolumn}
\usepackage{bm}
\begin{document}

\title{Extended \textit{versus} localized vibrations: the case of L-cysteine and L-cystine amino acids}
\author{M. S. Ishikawa}
\author{T. A. Lima}
\author{F. F. Ferreira}
\author{H. S. Martinho}
\email{herculano.martinho@ufabc.edu.br}

\affiliation{Centro de Ciências Naturais e Humanas, Universidade Federal do ABC, Santo André-SP, 09210-580, Brazil}

\begin{abstract}
A detailed quantitative analysis of the specific heat in the $1.8-300$ K temperature range for L-cysteine and L-cystine amino acids was  presented. We observed not extended but a sharp transition at $\sim 76$ K for L-cysteine. This transition was associated to the thiol group ordering and the order-disorder transition was adequately modeled by a 2D Ising model. The energy difference among two thiol configurations was found to be $-J=\varepsilon_{A}-\varepsilon_{B}=-66.6$ cal/mole. Besides, we conducted a study of phonon and rotor contributions to the specific heat and we proposed a generalization of Debye model. It was possible to evaluate the exponent of the $g(\omega)$, leading to the result that it corresponds to the Debye model for L-cysteine, which implies that the boson peak in this system is due to a maximum in the $C_{coup}(\omega)$ and also that the plane wave of wave-vector $\vec{q}$ is a good approximation to describe the phonons. On the other hand the origin of the boson peak for L-cystine correlates to a peak in $g(\omega)$ and phonons in L-cystine could be well represented by strongly attenuated plane waves or localized vibrations. Lastly, the analysis at very low temperature ($T<3$ K) indicated that L-cysteine presented a nearly temperature independent behaviour which is opposite to which is widely observed in systems with glassy characteristics within the Two-Level System (TLS) framework.
\end{abstract}

\maketitle

\section{Introduction}

It has been reported that biological macromolecules present two dynamical transitions at $T_{D}\sim200-230$ K and $T^{*}\sim80-100$ K \cite{frauenfelder88,doster89,parak89,roh05,roh06,chen06,frauenfelder09}. The first one occurs at hydration levels greater than $\sim18\%$, and it is related to a deviation from anharmonic to harmonic behavior of the mean squared atomic displacement with decrease of the temperature \cite{roh05}. According to some authors (see e.g. Ref. \cite{ciliberti06}) $T_{D}$ corresponds to the onset of a glass transition, $T_{g}$, although some researchers \cite{khodadadi09} pointed out that $T_{D}$ and $T_{g}$ have different physical origin. Additionally, it has been suggested that $T_{D}$ is correlated to onset of biochemical activities of the macromolecule \cite{frauenfelder88,parak89,doster89,roh05,berntsen05purple}.
 
Recent work \cite{lima14} reported that some physical properties of hydrated L-cysteine resemble those of quantum glass materials. Furthermore, a universal feature of such systems is that the vibrational density of state ($g(\omega$)) departs from the squared-frequency Debye-law, displaying an excess of states, the boson peak \cite{grigera03}. Transition at $T^{*}$ is hydration level independent \cite{roh05}. Some works \cite{roh05,schiro10} interpreted this transition as related to the thermal activation of methyl groups rotation. However, neither the microscopic nature nor the biological relation of these transitions is completely understood.

A detailed investigation of the crystal structures of amino acids and their dynamics is very important to understand complexes biological molecules \cite{boldyreva08book,kolesov08}. Besides, it was shown \cite{limat12,schiro11} that both transitions do not require the protein polypeptide chain as well as the protein secondary and tertiary structure. Intramolecular motions and intermolecular interactions could be probed by experimental techniques where temperature and pressure are tuning parameters \cite{kolesov08}.
 
Several studies based on the calorimetric measurements of amino acids that revealed phase transition can be mentioned. Wang \emph{et al.} \cite{wang94} observed a $\lambda-$transition at $272$ K by differential scanning calorimetry (DSC) for D-valine. It was proposed that the shape of the jump for D-valine is due to electron coupling. For taurine, Lima \emph{et al.} \cite{lima01} found the existence of a first-order transition at $251$ K with temperature-dependent Raman spectroscopy that was confirmed by DSC data. Besides, it was observed by Drebushchak \emph{et al.} \cite{drebushchak05} a second order phase transition near $252$ K for $\beta$ polymorph of glycine comparing the results with the data for $\alpha$-glycine. Such transition was considered as ferroelectric-paraelectric transition.

The orthorhombic polymorph of the amino acid L-cysteine has been also focus of recent interest. This amino acid possesses a very simple chemical structure and high biological relevance. The thiol or sulfurous group in the residues of L-cysteine is the most chemically reactive site in proteins under physiological conditions \cite{friedman1973chemistry}. This compound presents a tiny specific heat anomaly near $\sim76$ K \cite{paukov07,paukov08}. These authors presented a qualitative interpretation of this anomaly to an order-disorder phase transition bearing in mind the results of ref. \cite{moggach05} where it was shown that the thiol groups are ordered at $30$ K. Paukov \emph{et al.} \cite{paukov07,paukov08} measured the L-cysteine specific heat in pulse and continuous modes in the region of the anomaly, however no sharp order-disorder transition was observed. Kolesov \emph{et al.} \cite{kolesov08} utilizing variable-temperature polarized Raman spectroscopy verified the dynamic transition related to switching from $S-H\cdots S$ hydrogen bonds to the $S-H\cdots O$ contacts is not sharp, but is extended in a wide temperature range as observed by Paukov \emph{et al.} \cite{paukov07,paukov08}. Another qualitative propose concerning the nature of the transition near $\sim76$ K for L-cysteine relies to the rotation of $CH_{2}$ group \cite{limat12}.

In the present work a detailed quantitative analysis of the specific heat in the $1.8-300$ K temperature range for L-cysteine is presented. A comparison with L-cystine amino acid, was also performed. L-cystine is formed by two cysteine molecules linked via a disulfide bond which prohibit thiol ordering \cite{roux10}. 

\section{Materials and Methods}

\subsection{Samples}

The samples used were the commercial powder of orthorhombic crystalline L-cysteine and hexagonal crystalline L-cystine from Sigma-Aldrich (purity of $97$\% and $98$\%, respectively). According to X-ray diffraction data obtained with STOE STADI-P diffractometer and performing the Rietveld method analysis (using GSAS+EXPGUI software \cite{larson94,toby01}), it was possible to verify that the crystal structure of L-cysteine is orthorhombic as previously determined in ref. \cite{kerr75}. The space group is $P2_12_12_1$ with $Z=4$ and unit cell lattice parameters $a=8.11639(7)$ \AA, $b=12.17169(11)$ \AA, and $c=5.42266(4)$ \AA. The crystal structure of L-cystine was refined as belonging to  space group $P6_122$ (hexagonal) $Z=6$  with unit cell lattice parameters $a=b=5.42264(5)$  \AA, and $c=56.2908(5)$ \AA, in accordance with Ref. \cite{oughton59}.

\subsection{Calorimetric measurements}

The calorimetric measurements were performed in the Physical Properties Measurements System (PPMS) with Evercool-II\textregistered option from Quantum Design Inc. This system employs a thermal-relaxation calorimeter that determine the specific heat of the sample by measuring the thermal response to a change in heating conditions \cite{lashley03}.

\subsection{Theoretical models}

It has been pointed out that a superposition of complexes contributions need to be considered to explain the specific heat of L-cysteine. Due to the strong anharmonicity of the system and the glassy behavior, the usual low temperature Debye contribution to specific heat $c_{D}\propto T^{3}$ need to be revised. Moreover, the methylene group rotors and the order-disorder contributions need also to be taken into account.

\subsubsection{Generalized Debye model (GDM)}

Commonly, the approach used to describe the acoustic phonon contributions to the specific heat is the Debye model. However it has been shown that amino acids such as L-cysteine presents glass-like behavior, e.g. an excess contribution to the usual $g(\omega)$ that can be observed in the scaled specific heat $c_{p}(T)/T^3$ at low temperatures \cite{lima14}. Within the phonon localization picture model for boson peak \cite{elliott1992unified} one expects that the linear phonon dispersion law breaks down. We proposed a power-law for dispersion relation $\omega\left(q\right)=vq^{\alpha}$ where $v$ is the sound speed of a transverse or longitudinal phonon. Therefore, the $g(\omega)$ will be written as

\begin{equation}\label{g}
g(\omega)=\frac{3\mathcal{V}}{2\pi^2 \alpha v}\left( \frac{\omega}{v}\right)^{3/\alpha - 1},
\end{equation}

\noindent where $\mathcal{V}$ is the unit cell volume. The total internal energy is 

\begin{equation}\label{energy}
U=\frac{3\mathcal{V}\hbar}{2\pi^2 \alpha v^{3/
\alpha}}\int_{0}^{\infty}\frac{\omega^{3/\alpha}}{e^{\frac{\hbar\omega}{k_{B}T}}-1}d\omega,
\end{equation}

\noindent where $\hbar$ is the reduced Planck constant, $k_{B}$ is the Boltzmann constant. A sharp cutoff at $\omega_{c}$ is chosen that the total number of modes equals the number of vibrational degrees of freedom, $3\mathcal{N}$. Thus, 

\begin{equation}\label{cutoff}
\omega_{c}=v\left(\frac{6\pi^{2}\mathcal{N}}{\mathcal{V}}\right)^{\alpha/3}.
\end{equation}

The molar specific heat is calculated by

\begin{eqnarray}\label{phonon}
c_{\mathcal{V}}^{phonons}=\frac{N_{A}}{\mathcal{N}\mathcal{V}}\left(\frac{\partial U}{\partial T }\right)_{\mathcal{V}}= \nonumber\\
n\frac{9R}{\alpha}\left(\frac{T}{\theta_{c}}\right)^{3/\alpha}\int_{0}^{\theta_{c}/{T}}\frac{x^{3/\alpha+1}e^{x}}{\left(e^{x}-1\right)^2}dx.
\end{eqnarray}
\noindent where $R$ is the gas constant, $\hbar\omega_{c}=k_{B}\Theta_{c}$ and $n$ is the number of atoms per unit formula. In the case of the L-cysteine and the L-cystine $n=14$ and $28$, respectively. Note that the usual Debye model is the particular $\alpha=1$

\subsubsection{Specific heat of anisotropic rigid rotors}

Other relevant contribution to specific heat of biomolecules to be considered arises from the methyl or methylene groups rotations. Once approximating this rotating side chain as an anisotropic rigid body one could use the results of Caride and Tsallis \cite{caride84} and compute this contribution to the molar specific heat as

\begin{subequations} \label{rotors}
\begin{equation}\label{hcrotor}
c_{\mathcal{V}}^{rotors}=R\frac{1}{T^2}\left\{\frac{V}{Z}-\left(\frac{W}{Z}\right)^2\right\}
\end{equation}

with 

\begin{multline}\label{hcrotorv}
V\equiv\sum_{l=0}^{\infty} (2l+1)e^{\frac{-l(l+1)}{t}}\times\\
\sum_{m=-l}^{l} \left\{l(l+1)+\left(\frac{I_{xy}}{I_z}-1\right)m^2\right\}^2 e^{\frac{\frac{I_{xy}}{I_z}-1}{t}},
\end{multline}
\begin{multline}\label{hcrotorw}
W\equiv\sum_{l=0}^{\infty} (2l+1)e^{\frac{-l(l+1)}{t}}\times\\
\sum_{m=-l}^{l} \left\{l(l+1)+\left(\frac{I_{xy}}{I_z}-1\right)m^2\right\} e^{\frac{\frac{I_{xy}}{I_z}-1}{t}},
\end{multline}
\begin{equation}\label{hcrotort}
Z\equiv\sum_{l=0}^{\infty} (2l+1)e^{\frac{-l(l+1)}{t}}\sum_{m=-l}^{l} e^{\frac{\frac{I_{xy}}{I_z}-1}{t}},
\end{equation}
\end{subequations}

\noindent where $l$ is the angular momentum, $I_x=I_y=I_{xy}$ is the moment of inertia about $x$ and $y$ axes with the same module, $I_z$ the moment of inertia about $z$ axis, $m$ is the magnetic quantum number ranging from $-l,-l+1,\dots,l$.

\subsubsection{Order-disorder and Ising model}

Once making the analogy of the two possible states of the spins with the two possible states of the thiol groups in the plane the hypothesis of order-disorder transition at $\sim 76$ K for L-cysteine could be modeled by using the 2D Ising model \cite{onsager44}. Thereby, the molar specific heat at the order-disorder transition is given by

\begin{widetext}
\begin{eqnarray}\label{ising}
c_{\mathcal{V}}^{Ising}(T)=\frac{2R}{\pi} K^{2}\coth^2(2K)\left\{2K(\kappa)-2E(\kappa)-2\sech^2(2K)\left(\frac{\pi}{2}+(2\tanh(2K)-1)K(\kappa)\right)\right\},
\end{eqnarray}
\end{widetext}

\noindent with $\kappa=2\sinh(2K)/\cosh^2(2K)$, $K=J/k_{B}T$, $K(\kappa)$, and $E(\kappa)$ are elliptic integrals of the first and second type, respectively. The parameter $J$ corresponds to the average energy between the two average ordered and disordered structures.

\section{Results and Discussion}

The first question that will be addressed concerns the nature of the transition at $\sim 76$ K. As pointed by Lashley \textit{et al.} \cite{lashley03}, it is important to stress that there are many difficulties in analyzing the individual contributions from various degrees of freedom to the specific heat at low temperatures. Consequently, there is a great deal of effort that goes into controlling the details that are part of calorimetric measurements. Usually the control of details as thermometry, temperature-scale issues, and the creation and control of heat leaks specially in sharper transitions of first order is needed. In a commercially available calorimeter a large number of these details might be hidden from the user \cite{lashley03}. The specific heat can be determined in the vicinity of a phase transition by analyzing the thermal-relaxation data point-by-point rather than by obtaining a single $C_{\mathcal{P}}$ value for the entire temperature region spanned by the decay \cite{lashley03}. From the time-dependent relaxation data $T(t)$ the specific heat near a phase transition is obtained by

\begin{equation}\label{relax}
C_{\mathcal{P}}[T(t)]=-K\frac{(T-T_{0})}{dT(t)/dt},
\end{equation}
\noindent where $K$ is the thermal conductivity of the calorimeter wires and $T_{0}$ is thermal bath temperature.

\begin{figure}[tbh!]
\includegraphics[width=8.2cm,height=6cm]{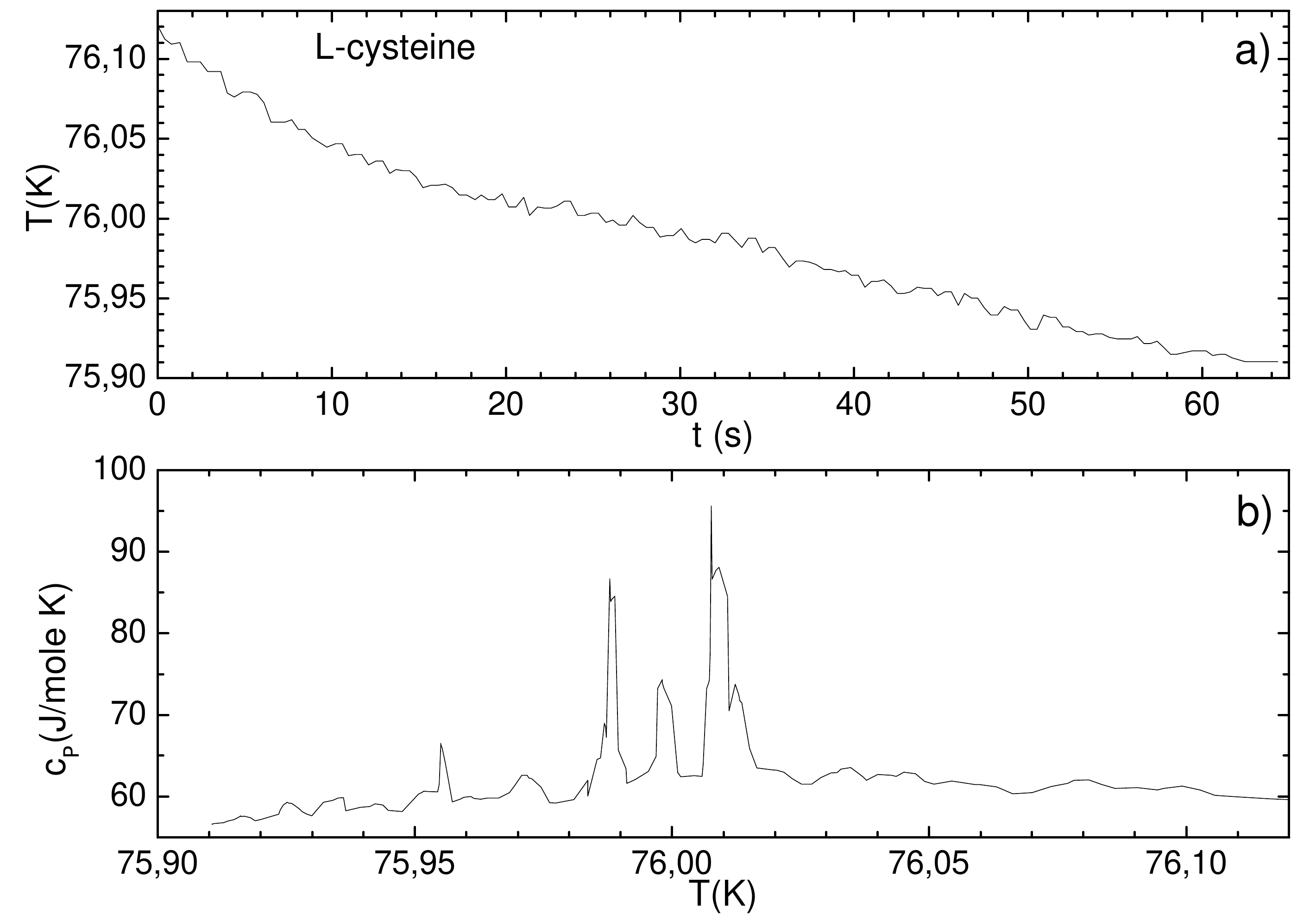}
\caption{a) Relaxation data used to determine $C_{p}(T)$ around $T_{c}\sim 76$ K. b) The specific heat data calculated in the vicinity of the first-order transition.} \label{transition}
\end{figure}

\begin{figure}[tbh!]
\includegraphics[height=10 cm]{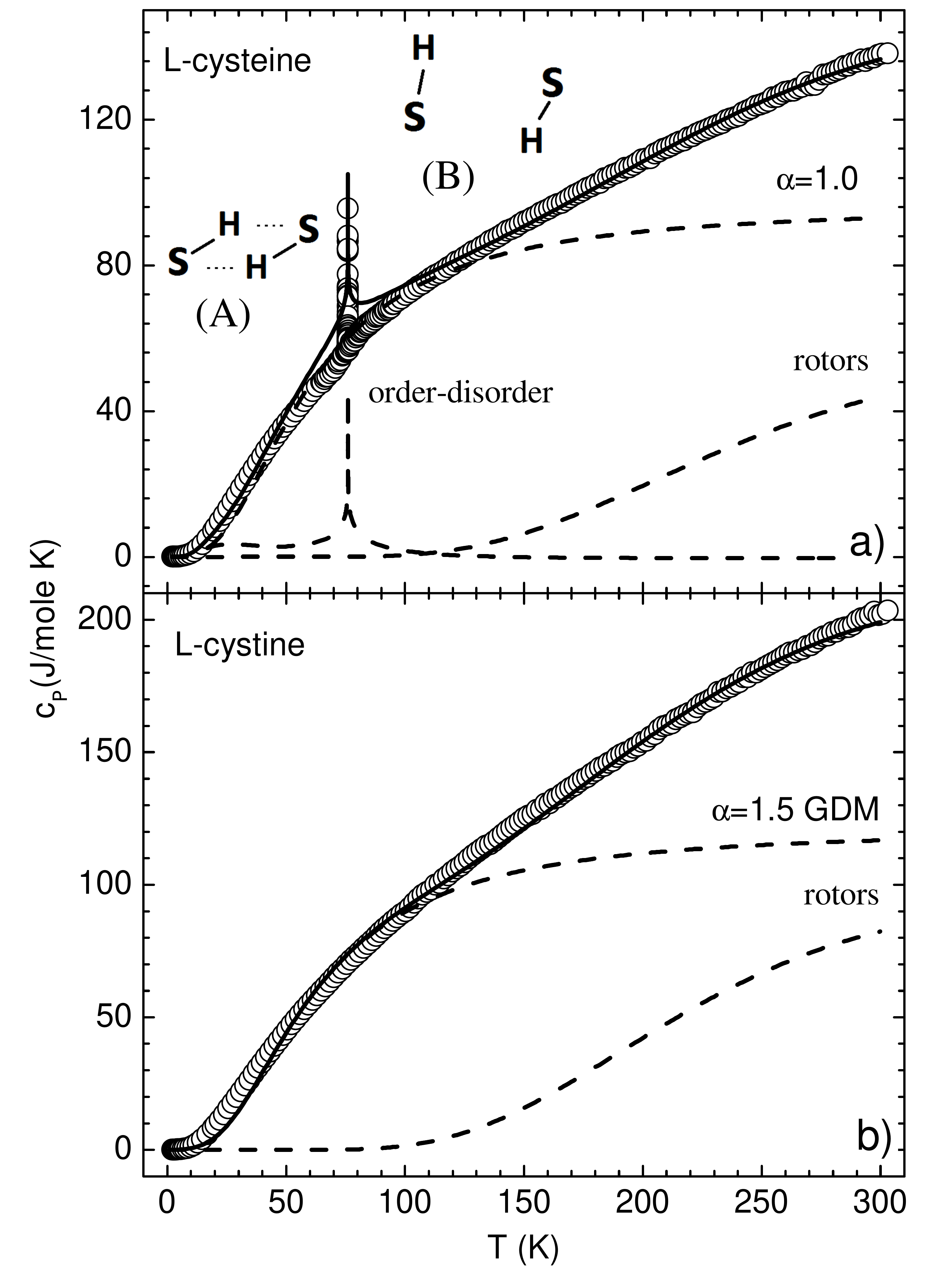}
\caption{Experimental data and best simulated curve using Eq. \ref{cptotal} for L-cysteine (a) and L-cystine (b). The Ising contribution was not considered for L-cystine.}
\label{simulations}
\end{figure}

Figure \ref{transition}a) presents the relaxation data $T(t)$ around $76$ K. The $C_{\mathcal{P}}[T(t)]$ data obtained by using Eq. \ref{relax} is shown on Fig. \ref{transition}b). In the narrow temperature window of $\sim 0.20$ K the sharpness of the transition becomes clear. The transition jump starts to develops at $T\sim75.96$ K ending at $T\sim76.96$ K with maximum of $\sim100$ J/mole K which is limited by experimental resolution. The jump in the specific heat compared to the overall $1.8-300$ K data is shown on Fig. \ref{simulations}a).  Figure \ref{simulations}b) shows the molar specific heat data for L-cystine amino acid. The absence of phase transition in this case corroborates the hypothesis that thiol order-disorder transition is responsible for the sharp peak observed at $76$ K for L-cysteine.  The ordered configuration corresponds to the (A) scheme shown on Fig. \ref{simulations}. The Ising model was able to reproduce the peak corresponding to ordering of thiol groups with acceptable accordance. The jump observed experimentally appeared to be narrow than the Ising simulation peak. This fact could be explained remembering that Eq. \ref{ising} does not take into the several possible thermally activated disordered configurations (B). Usually, Ising Monte Carlo simulations sampling several configurational possibilities furnish better accordance with experimental data. The energy cost of the thiol ordering is $-J=\varepsilon_{A}-\varepsilon_{B}=-66.6$ cal/mole.

Besides the ordering transition, contributions from phonons and rotors also need be considered. The quantity experimentally accessed by our experiments is $C_{\mathcal{P}}$. Since for solids $C_{\mathcal{V}}\approx C_{\mathcal{P}}$ and based on Eqs. \ref{phonon}, \ref{rotors}, \ref{ising} the total molar specific heat of L-cysteine could be modeled according to 

\begin{eqnarray}\label{cptotal}
c_{\mathcal{P}}=ac^{phonons}(\Theta_{c},\alpha,T)+c^{Ising}(J,T)\nonumber\\
+bc^{rotors}(L, I_{xy}, I_{z}, T),
\end{eqnarray}

\begin{figure}[tbh!]
\includegraphics[width=8cm]{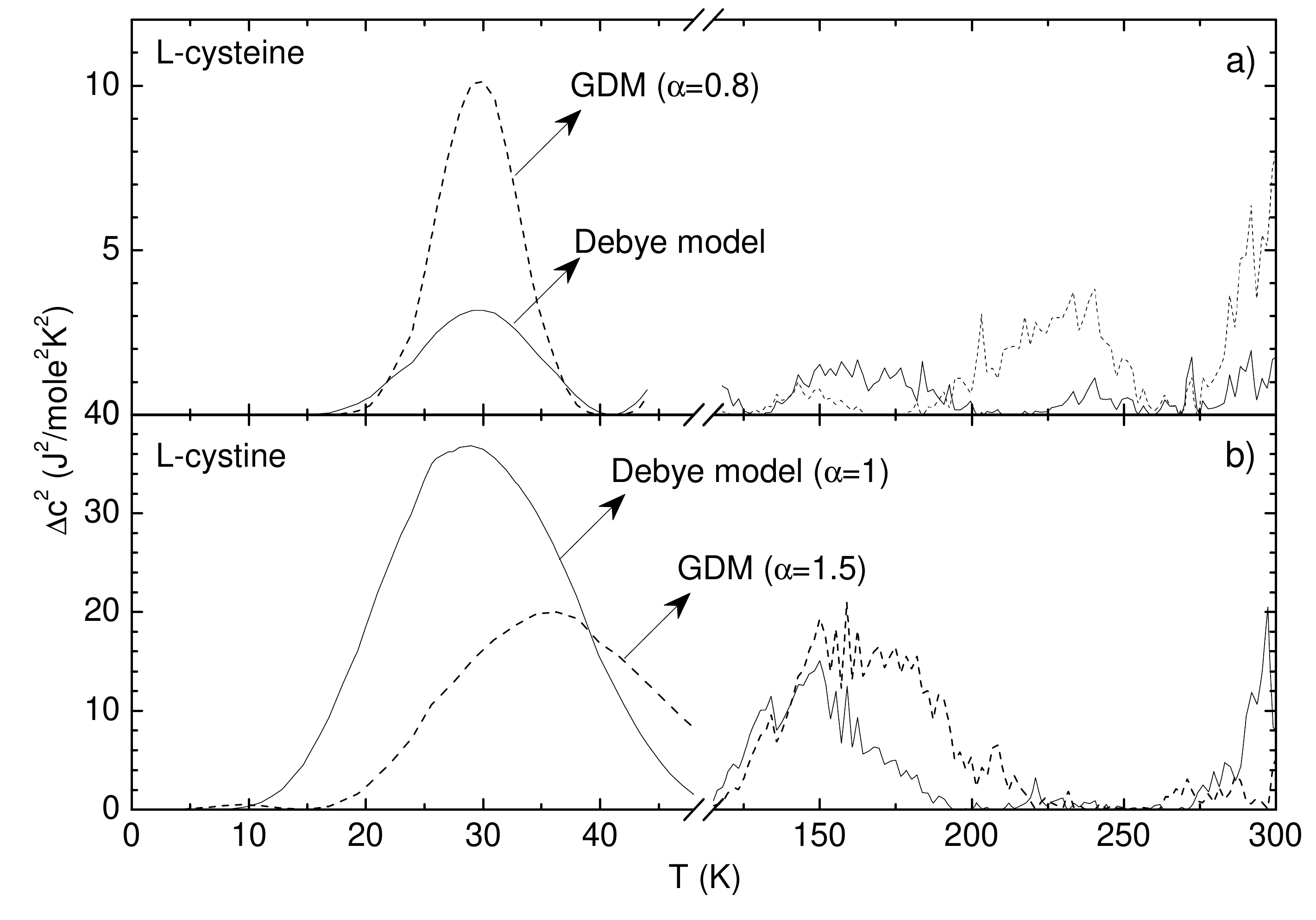}
\caption{Squared difference plot $\Delta^2C$ for L-cysteine (a) and L-cystine for some selected $\alpha$ values.}\label{compara}
\end{figure}

For the L-cystine case the contribution due to order-disorder of the thiol group was not considered. Selected simulations and each contribution to specific heat are shown in Fig. \ref{simulations} for L-cysteine and L-cystine. For both amino acids, it was performed simulations taking $\alpha=1$ (Debye model) and $\alpha \neq 1$. The goodness of simulations was evaluated by computing the difference $\Delta^2 c=(c_{experimental}-c_{simulated})^2$ (Fig. \ref{compara}).

Table \ref{table1} summarizes all obtained parameters. Moments of inertia of $CH_{2}$ are in accordance to prolate symmetric top ($I_{xy}<I_z$), which is consistent with the result found by Lima \textit{et al.} \cite{limat12}. $I_{xy}$ and $I_{z}$ for L-cystine were more sensible to $\alpha$ presenting $\sim 11$\% of variation. The obtained values are of the same magnitude order of those calculated from atom masses and distances.

\begin{table}[tbh!]
\caption{\label{table1}Table of the parameters obtained by simulation for L-cysteine and L-cystine.}
\begin{ruledtabular}
\begin{tabular}{ccccc}
 Parameters &\multicolumn{2}{c}{L-cysteine}&\multicolumn{2}{c}{L-cystine}\\ 
 \hline
 $\alpha$ 													& 1 		& 0.8 	& 1 		& 1.5 \\ 	
$a$																	& 0.275 & 0.265 & 0.173 & 0.38\\ 
$L [J\cdot s]$ 											& 10 		& 10 		& 10 		& 10 	\\ 
$I_{xy} [10^{-49} Kg \cdot m^{2}]$ 	& 3.95  & 4.3   & 4.25 	& 3.8 \\ 
$I_{z} [10^{-47} Kg \cdot m^2]$ 		& 9.55  & 9    	& 17 		& 10 	\\
$b$																	& 2.55 	& 2.45 	& 4.43 	& 4.7 \\ 
$\Theta_{c} [K]$ 										& 245 	& 225 	& 252 	& 320 \\ 
$J/k_{B}[K]$												& 33.5  & 33.5 	& -		 	& -		\\ 
\end{tabular}
\end{ruledtabular}
\end{table}

Phonon contributions to the simulations need to be analyzed in more details due to implications on the $g(\omega)$. It is the main contribution to low temperature $c_{\mathcal{P}}$. Confining our analysis to $T\leq 50$ K, one could conclude from Fig. \ref{compara} that L-cysteine data were best simulated with Debye model ($\alpha=1$). On the other hand, the $\alpha=1.5$ gave the best results for L-cystine. From values of $\alpha$ obtained by simulations we could infer that $g(\omega) \propto \omega^{2}$ ($\alpha=1$) and $g(\omega) \propto \omega^{2.75}$ ($\alpha=1.5$) for L-cysteine and L-cystine, respectively. These findings have very important implications to the boson peak origin comprehension.

Shuker \textit{et al.} \cite{shuker1970raman} have shown that the low frequency Raman scattering intensity of amorphous solid is

\begin{equation}
I(\omega, T)=C_{coup}(\omega)\frac{g(\omega)}{\omega}\left[1+n(\omega,T)\right],
\end{equation}

\noindent where $C_{coup}(\omega)$ is the coupling constant, $n(\omega,T)$ is the Bose-Einstein occupation factor. This expression have been widely used to explain the boson peak in glasses. 

Since $g(\omega)$ corresponds to the Debye model for L-cysteine, one could infer that the boson peak in this system does not have its origin due to a peak in the vibrational density of states but due a maximum in the $C_{coup}(\omega)$. Thereby, for L-cysteine the dispersion relation $\omega=vq$ is expected to be valid up to higher frequencies. Thus the plane wave of wave-vector $\vec{q}$ is a good approximation to describe the phonons \cite{ahmad1986}. However, for L-cystine, the excess of vibrational density of states compared to Debye model is clear from the exponent dependence of $g(\omega)$ and the origin of the boson peak for L-cystine correlates to a peak in $g(\omega)$. Therefore, phonons in L-cystine could be well represented by strongly attenuated plane waves or localized vibrations. This very distinct behavior has direct impact on $\Theta_{c}$ estimation since the localized vibrations results an increase of $75$ K for this parameter.

\begin{figure}[tbh!]
\includegraphics[width=7cm,height=4.8cm]{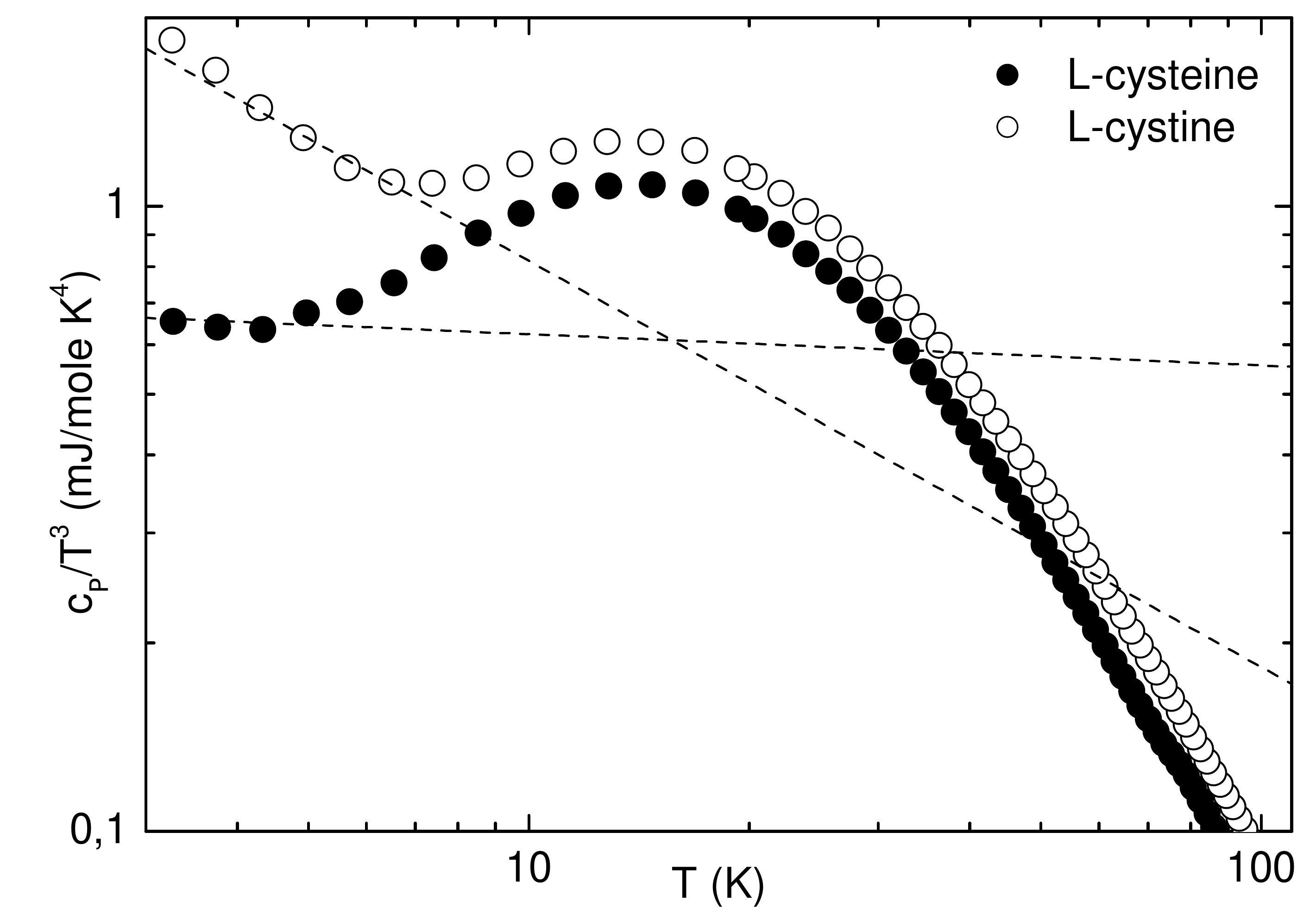}
\caption{$c_{p}/T^{3}$ \textit{vs} $T$ specific heat data of L-cysteine (closed circles) and L-cystine (open circles). Dashed lines represent the TLS contribution $\varpropto T^{-0.005}$ and $\varpropto T^{-0.65}$ for L-cysteine and L-cystine, respectively.}\label{TLScontribution}
\end{figure}

At very low temperatures ($\sim 1$ K) the specific heat of glasses is usually described in the two-level systems (TLS) model framework \cite{binder}. This model assumes the glass state as ``frozen liquid", with a large number of metastable states. A very low temperatures thermally activated processes between these states are highly improbable. Therefore one is left with the idea of tunneling process between two states that correspondS to the two local minima of configuration \cite{binder}. Fig. \ref{TLScontribution} shows the log-log plot of $c_{p}/T^{3}$ \textit{vs} $T$ for L-cysteine and L-cystine samples. The Debye plus TLS contribution was fitted to $\varpropto T^{-0.005}$ and $\varpropto T^{-0.65}$ for L-cysteine and L-cystine, respectively.

\section{Conclusion}

From our quantitative analysis of the specific heat results for L-cysteine and L-cystine we conclude that the transition at $\sim 76$ K for L-cysteine is due to thiol group ordering. We show that this transition is not extended as presented in literature, but is a sharp first order phase transition. We elaborate that its sharpness prevented others researchers to clearly observe the emergence of the peak. The order-disorder transition was adequately modeled by Ising model. The energy cost of the thiol ordering was obtained as $-J=\varepsilon_{A}-\varepsilon_{B}= - 66.6$ cal/mole. Phonon and rotor contributions were also analyzed. From the conjugated analysis it was possible estimate the exponent of the $g(\omega)$. It was found that it corresponds to the Debye model for L-cysteine, which imply that the boson peak in this system is due to a maximum in the $C_{coup}(\omega)$ and also that the plane wave of wave-vector $\vec{q}$ is a good approximation to describe the phonons. On the other hand, for L-cystine, the origin of the boson peak correlates to a peak in $g(\omega)$ and phonons in L-cystine could be well represented by strongly attenuated plane waves or localized vibrations. Analysis at very low temperature ($T<1$ K) indicates that L-cysteine presented a nearly temperature independent behavior of is a remarkable finding for a system with glass characteristics since does not follow the prevision of TLS model.

\begin{acknowledgments}
The authors would like to thank the Brazilian agencies CNPq and FAPESP for their financial support and the Multiuser Central Facilities at UFABC (CEM-UFABC) for providing conditions to perform the experiments described in this work.
\end{acknowledgments}

\bibliography{ref}

\begin{thebibliography}{36}%
\makeatletter
\providecommand \@ifxundefined [1]{%
 \@ifx{#1\undefined}
}%
\providecommand \@ifnum [1]{%
 \ifnum #1\expandafter \@firstoftwo
 \else \expandafter \@secondoftwo
 \fi
}%
\providecommand \@ifx [1]{%
 \ifx #1\expandafter \@firstoftwo
 \else \expandafter \@secondoftwo
 \fi
}%
\providecommand \natexlab [1]{#1}%
\providecommand \enquote  [1]{``#1''}%
\providecommand \bibnamefont  [1]{#1}%
\providecommand \bibfnamefont [1]{#1}%
\providecommand \citenamefont [1]{#1}%
\providecommand \href@noop [0]{\@secondoftwo}%
\providecommand \href [0]{\begingroup \@sanitize@url \@href}%
\providecommand \@href[1]{\@@startlink{#1}\@@href}%
\providecommand \@@href[1]{\endgroup#1\@@endlink}%
\providecommand \@sanitize@url [0]{\catcode `\\12\catcode `\$12\catcode
  `\&12\catcode `\#12\catcode `\^12\catcode `\_12\catcode `\%12\relax}%
\providecommand \@@startlink[1]{}%
\providecommand \@@endlink[0]{}%
\providecommand \url  [0]{\begingroup\@sanitize@url \@url }%
\providecommand \@url [1]{\endgroup\@href {#1}{\urlprefix }}%
\providecommand \urlprefix  [0]{URL }%
\providecommand \Eprint [0]{\href }%
\providecommand \doibase [0]{http://dx.doi.org/}%
\providecommand \selectlanguage [0]{\@gobble}%
\providecommand \bibinfo  [0]{\@secondoftwo}%
\providecommand \bibfield  [0]{\@secondoftwo}%
\providecommand \translation [1]{[#1]}%
\providecommand \BibitemOpen [0]{}%
\providecommand \bibitemStop [0]{}%
\providecommand \bibitemNoStop [0]{.\EOS\space}%
\providecommand \EOS [0]{\spacefactor3000\relax}%
\providecommand \BibitemShut  [1]{\csname bibitem#1\endcsname}%
\let\auto@bib@innerbib\@empty
\bibitem [{\citenamefont {Frauenfelder}\ \emph {et~al.}(1988)\citenamefont
  {Frauenfelder}, \citenamefont {Parak},\ and\ \citenamefont
  {Young}}]{frauenfelder88}%
  \BibitemOpen
  \bibfield  {author} {\bibinfo {author} {\bibfnamefont {H.}~\bibnamefont
  {Frauenfelder}}, \bibinfo {author} {\bibfnamefont {F.}~\bibnamefont {Parak}},
  \ and\ \bibinfo {author} {\bibfnamefont {R.~D.}\ \bibnamefont {Young}},\
  }\href@noop {} {\bibfield  {journal} {\bibinfo  {journal} {Ann. Rev. Biophys.
  Biophys. Chem.}\ }\textbf {\bibinfo {volume} {17}},\ \bibinfo {pages} {451}
  (\bibinfo {year} {1988})}\BibitemShut {NoStop}%
\bibitem [{\citenamefont {Doster}\ \emph {et~al.}(1989)\citenamefont {Doster},
  \citenamefont {Cusack},\ and\ \citenamefont {Petry}}]{doster89}%
  \BibitemOpen
  \bibfield  {author} {\bibinfo {author} {\bibfnamefont {W.}~\bibnamefont
  {Doster}}, \bibinfo {author} {\bibfnamefont {S.}~\bibnamefont {Cusack}}, \
  and\ \bibinfo {author} {\bibfnamefont {W.}~\bibnamefont {Petry}},\
  }\href@noop {} {\bibfield  {journal} {\bibinfo  {journal} {Nature}\ }\textbf
  {\bibinfo {volume} {337}},\ \bibinfo {pages} {754} (\bibinfo {year}
  {1989})}\BibitemShut {NoStop}%
\bibitem [{\citenamefont {Parak}\ and\ \citenamefont {Knapp}(1984)}]{parak89}%
  \BibitemOpen
  \bibfield  {author} {\bibinfo {author} {\bibfnamefont {F.}~\bibnamefont
  {Parak}}\ and\ \bibinfo {author} {\bibfnamefont {E.~W.}\ \bibnamefont
  {Knapp}},\ }\href {\doibase 10.1073/pnas.81.22.7088} {\bibfield  {journal}
  {\bibinfo  {journal} {Proc. Natl. Acad. Sci. (U.S.A.)}\ }\textbf {\bibinfo
  {volume} {81}},\ \bibinfo {pages} {7088} (\bibinfo {year}
  {1984})}\BibitemShut {NoStop}%
\bibitem [{\citenamefont {Roh}\ \emph {et~al.}(2005)\citenamefont {Roh},
  \citenamefont {Novikov}, \citenamefont {Gregory}, \citenamefont {Curtis},
  \citenamefont {Chowdhuri},\ and\ \citenamefont {Sokolov}}]{roh05}%
  \BibitemOpen
  \bibfield  {author} {\bibinfo {author} {\bibfnamefont {J.~H.}\ \bibnamefont
  {Roh}}, \bibinfo {author} {\bibfnamefont {V.~N.}\ \bibnamefont {Novikov}},
  \bibinfo {author} {\bibfnamefont {R.~B.}\ \bibnamefont {Gregory}}, \bibinfo
  {author} {\bibfnamefont {J.~E.}\ \bibnamefont {Curtis}}, \bibinfo {author}
  {\bibfnamefont {Z.}~\bibnamefont {Chowdhuri}}, \ and\ \bibinfo {author}
  {\bibfnamefont {A.~P.}\ \bibnamefont {Sokolov}},\ }\href@noop {} {\bibfield
  {journal} {\bibinfo  {journal} {Phys. Rev. Lett.}\ }\textbf {\bibinfo
  {volume} {95}} (\bibinfo {year} {2005})}\BibitemShut {NoStop}%
\bibitem [{\citenamefont {Roh}\ \emph {et~al.}(2006)\citenamefont {Roh},
  \citenamefont {Curtis}, \citenamefont {Azzam}, \citenamefont {Novikov},
  \citenamefont {Peral}, \citenamefont {Chowdhuri}, \citenamefont {Gregory},\
  and\ \citenamefont {Sokolov}}]{roh06}%
  \BibitemOpen
  \bibfield  {author} {\bibinfo {author} {\bibfnamefont {J.~H.}\ \bibnamefont
  {Roh}}, \bibinfo {author} {\bibfnamefont {J.~E.}\ \bibnamefont {Curtis}},
  \bibinfo {author} {\bibfnamefont {S.}~\bibnamefont {Azzam}}, \bibinfo
  {author} {\bibfnamefont {V.~N.}\ \bibnamefont {Novikov}}, \bibinfo {author}
  {\bibfnamefont {I.}~\bibnamefont {Peral}}, \bibinfo {author} {\bibfnamefont
  {Z.}~\bibnamefont {Chowdhuri}}, \bibinfo {author} {\bibfnamefont {R.~B.}\
  \bibnamefont {Gregory}}, \ and\ \bibinfo {author} {\bibfnamefont {A.~P.}\
  \bibnamefont {Sokolov}},\ }\href {\doibase 10.1529/biophysj.106.082214}
  {\bibfield  {journal} {\bibinfo  {journal} {Biophys. J.}\ }\textbf {\bibinfo
  {volume} {91}},\ \bibinfo {pages} {2573} (\bibinfo {year}
  {2006})}\BibitemShut {NoStop}%
\bibitem [{\citenamefont {Chen}\ \emph {et~al.}(2006)\citenamefont {Chen},
  \citenamefont {Liu}, \citenamefont {Fratini}, \citenamefont {Baglioni},
  \citenamefont {Faraone},\ and\ \citenamefont {Mamontov}}]{chen06}%
  \BibitemOpen
  \bibfield  {author} {\bibinfo {author} {\bibfnamefont {S.~H.}\ \bibnamefont
  {Chen}}, \bibinfo {author} {\bibfnamefont {L.}~\bibnamefont {Liu}}, \bibinfo
  {author} {\bibfnamefont {E.}~\bibnamefont {Fratini}}, \bibinfo {author}
  {\bibfnamefont {P.}~\bibnamefont {Baglioni}}, \bibinfo {author}
  {\bibfnamefont {A.}~\bibnamefont {Faraone}}, \ and\ \bibinfo {author}
  {\bibfnamefont {E.}~\bibnamefont {Mamontov}},\ }\href {\doibase
  10.1073/pnas.0602474103} {\bibfield  {journal} {\bibinfo  {journal} {Proc.
  Natl. Acad. Sci. (U.S.A.)}\ }\textbf {\bibinfo {volume} {103}},\ \bibinfo
  {pages} {9012} (\bibinfo {year} {2006})}\BibitemShut {NoStop}%
\bibitem [{\citenamefont {Frauenfelder}\ \emph {et~al.}(2009)\citenamefont
  {Frauenfelder}, \citenamefont {Chen}, \citenamefont {Berendzen},
  \citenamefont {Fenimore}, \citenamefont {Jansson}, \citenamefont {McMahon},
  \citenamefont {Stroe}, \citenamefont {Swenson},\ and\ \citenamefont
  {Young}}]{frauenfelder09}%
  \BibitemOpen
  \bibfield  {author} {\bibinfo {author} {\bibfnamefont {H.}~\bibnamefont
  {Frauenfelder}}, \bibinfo {author} {\bibfnamefont {G.}~\bibnamefont {Chen}},
  \bibinfo {author} {\bibfnamefont {J.}~\bibnamefont {Berendzen}}, \bibinfo
  {author} {\bibfnamefont {P.~W.}\ \bibnamefont {Fenimore}}, \bibinfo {author}
  {\bibfnamefont {H.}~\bibnamefont {Jansson}}, \bibinfo {author} {\bibfnamefont
  {B.~H.}\ \bibnamefont {McMahon}}, \bibinfo {author} {\bibfnamefont {I.~R.}\
  \bibnamefont {Stroe}}, \bibinfo {author} {\bibfnamefont {J.}~\bibnamefont
  {Swenson}}, \ and\ \bibinfo {author} {\bibfnamefont {R.~D.}\ \bibnamefont
  {Young}},\ }\href {\doibase 10.1073/pnas.0900336106} {\bibfield  {journal}
  {\bibinfo  {journal} {Proc. Natl. Acad. Sci. (U.S.A.)}\ }\textbf {\bibinfo
  {volume} {106}},\ \bibinfo {pages} {5129} (\bibinfo {year}
  {2009})}\BibitemShut {NoStop}%
\bibitem [{\citenamefont {Ciliberti}\ \emph {et~al.}(2006)\citenamefont
  {Ciliberti}, \citenamefont {De~Los~Rios},\ and\ \citenamefont
  {Piazza}}]{ciliberti06}%
  \BibitemOpen
  \bibfield  {author} {\bibinfo {author} {\bibfnamefont {S.}~\bibnamefont
  {Ciliberti}}, \bibinfo {author} {\bibfnamefont {P.}~\bibnamefont
  {De~Los~Rios}}, \ and\ \bibinfo {author} {\bibfnamefont {F.}~\bibnamefont
  {Piazza}},\ }\href@noop {} {\bibfield  {journal} {\bibinfo  {journal} {Phys.
  Rev. Lett.}\ }\textbf {\bibinfo {volume} {96}} (\bibinfo {year}
  {2006})}\BibitemShut {NoStop}%
\bibitem [{\citenamefont {Khodadadi}\ \emph {et~al.}(2010)\citenamefont
  {Khodadadi}, \citenamefont {Malkovskiy}, \citenamefont {Kisliuk},\ and\
  \citenamefont {Sokolov}}]{khodadadi09}%
  \BibitemOpen
  \bibfield  {author} {\bibinfo {author} {\bibfnamefont {S.}~\bibnamefont
  {Khodadadi}}, \bibinfo {author} {\bibfnamefont {A.}~\bibnamefont
  {Malkovskiy}}, \bibinfo {author} {\bibfnamefont {A.}~\bibnamefont {Kisliuk}},
  \ and\ \bibinfo {author} {\bibfnamefont {A.}~\bibnamefont {Sokolov}},\ }\href
  {\doibase http://dx.doi.org/10.1016/j.bbapap.2009.05.006} {\bibfield
  {journal} {\bibinfo  {journal} {Biochim. Biophys. Acta.}\ }\textbf {\bibinfo
  {volume} {1804}},\ \bibinfo {pages} {15 } (\bibinfo {year}
  {2010})}\BibitemShut {NoStop}%
\bibitem [{ber(2005)}]{berntsen05purple}%
  \BibitemOpen
  \href@noop {} {\bibfield  {journal} {\bibinfo  {journal} {Biophys. J.}\
  }\textbf {\bibinfo {volume} {89}},\ \bibinfo {pages} {3120 } (\bibinfo {year}
  {2005})}\BibitemShut {NoStop}%
\bibitem [{\citenamefont {Lima}\ \emph {et~al.}(2013)\citenamefont {Lima},
  \citenamefont {Ishikawa},\ and\ \citenamefont {Martinho}}]{lima14}%
  \BibitemOpen
  \bibfield  {author} {\bibinfo {author} {\bibfnamefont {T.~A.}\ \bibnamefont
  {Lima}}, \bibinfo {author} {\bibfnamefont {M.~S.}\ \bibnamefont {Ishikawa}},
  \ and\ \bibinfo {author} {\bibfnamefont {H.~S.}\ \bibnamefont {Martinho}},\
  }\href@noop {} {\bibfield  {journal} {\bibinfo  {journal} {arXiv preprint
  arXiv:1309.0412}\ } (\bibinfo {year} {2013})}\BibitemShut {NoStop}%
\bibitem [{\citenamefont {Grigera}\ \emph {et~al.}(2003)\citenamefont
  {Grigera}, \citenamefont {Martin-Mayor}, \citenamefont {Parisi},\ and\
  \citenamefont {Verrocchio}}]{grigera03}%
  \BibitemOpen
  \bibfield  {author} {\bibinfo {author} {\bibfnamefont {T.~S.}\ \bibnamefont
  {Grigera}}, \bibinfo {author} {\bibfnamefont {V.}~\bibnamefont
  {Martin-Mayor}}, \bibinfo {author} {\bibfnamefont {G.}~\bibnamefont
  {Parisi}}, \ and\ \bibinfo {author} {\bibfnamefont {P.}~\bibnamefont
  {Verrocchio}},\ }\href {\doibase 10.1038/nature01475} {\bibfield  {journal}
  {\bibinfo  {journal} {Nature}\ }\textbf {\bibinfo {volume} {422}},\ \bibinfo
  {pages} {289} (\bibinfo {year} {2003})}\BibitemShut {NoStop}%
\bibitem [{\citenamefont {Schir{\'o}}\ \emph {et~al.}(2010)\citenamefont
  {Schir{\'o}}, \citenamefont {Caronna}, \citenamefont {Natali},\ and\
  \citenamefont {Cupane}}]{schiro10}%
  \BibitemOpen
  \bibfield  {author} {\bibinfo {author} {\bibfnamefont {G.}~\bibnamefont
  {Schir{\'o}}}, \bibinfo {author} {\bibfnamefont {C.}~\bibnamefont {Caronna}},
  \bibinfo {author} {\bibfnamefont {F.}~\bibnamefont {Natali}}, \ and\ \bibinfo
  {author} {\bibfnamefont {A.}~\bibnamefont {Cupane}},\ }\href@noop {}
  {\bibfield  {journal} {\bibinfo  {journal} {Phys. Chem. Chem. Phys.}\
  }\textbf {\bibinfo {volume} {12}},\ \bibinfo {pages} {10215} (\bibinfo {year}
  {2010})}\BibitemShut {NoStop}%
\bibitem [{\citenamefont {Boldyreva}(2008)}]{boldyreva08book}%
  \BibitemOpen
  \bibfield  {author} {\bibinfo {author} {\bibfnamefont {E.}~\bibnamefont
  {Boldyreva}},\ }in\ \href {\doibase 10.1007/978-1-4020-5941-4_7} {\emph
  {\bibinfo {booktitle} {Models, Mysteries and Magic of Molecules}}},\ \bibinfo
  {editor} {edited by\ \bibinfo {editor} {\bibfnamefont {J.~C.~A.}\
  \bibnamefont {Boeyens}}\ and\ \bibinfo {editor} {\bibfnamefont {J.~F.}\
  \bibnamefont {Ogilvie}}}\ (\bibinfo  {publisher} {Springer Netherlands},\
  \bibinfo {year} {2008})\ pp.\ \bibinfo {pages} {167--192}\BibitemShut
  {NoStop}%
\bibitem [{\citenamefont {Kolesov}\ \emph {et~al.}(2008)\citenamefont
  {Kolesov}, \citenamefont {Minkov}, \citenamefont {Boldyreva},\ and\
  \citenamefont {Drebushchak}}]{kolesov08}%
  \BibitemOpen
  \bibfield  {author} {\bibinfo {author} {\bibfnamefont {B.~A.}\ \bibnamefont
  {Kolesov}}, \bibinfo {author} {\bibfnamefont {V.~S.}\ \bibnamefont {Minkov}},
  \bibinfo {author} {\bibfnamefont {E.~V.}\ \bibnamefont {Boldyreva}}, \ and\
  \bibinfo {author} {\bibfnamefont {T.~N.}\ \bibnamefont {Drebushchak}},\
  }\href {\doibase 10.1021/jp804142c} {\bibfield  {journal} {\bibinfo
  {journal} {J. Phys. Chem. B}\ }\textbf {\bibinfo {volume} {112}},\ \bibinfo
  {pages} {12827} (\bibinfo {year} {2008})}\BibitemShut {NoStop}%
\bibitem [{\citenamefont {Lima}\ \emph {et~al.}(2012)\citenamefont {Lima},
  \citenamefont {Sato}, \citenamefont {Martins}, \citenamefont {Homem-de
  Mello}, \citenamefont {Lago}, \citenamefont {Coutinho-Neto}, \citenamefont
  {Ferreira}, \citenamefont {Giles}, \citenamefont {Pires},\ and\ \citenamefont
  {Martinho}}]{limat12}%
  \BibitemOpen
  \bibfield  {author} {\bibinfo {author} {\bibfnamefont {T.~A.}\ \bibnamefont
  {Lima}}, \bibinfo {author} {\bibfnamefont {E.~T.}\ \bibnamefont {Sato}},
  \bibinfo {author} {\bibfnamefont {E.~T.}\ \bibnamefont {Martins}}, \bibinfo
  {author} {\bibfnamefont {P.}~\bibnamefont {Homem-de Mello}}, \bibinfo
  {author} {\bibfnamefont {A.~F.}\ \bibnamefont {Lago}}, \bibinfo {author}
  {\bibfnamefont {M.~D.}\ \bibnamefont {Coutinho-Neto}}, \bibinfo {author}
  {\bibfnamefont {F.~F.}\ \bibnamefont {Ferreira}}, \bibinfo {author}
  {\bibfnamefont {C.}~\bibnamefont {Giles}}, \bibinfo {author} {\bibfnamefont
  {M.~O.~C.}\ \bibnamefont {Pires}}, \ and\ \bibinfo {author} {\bibfnamefont
  {H.}~\bibnamefont {Martinho}},\ }\href@noop {} {\bibfield  {journal}
  {\bibinfo  {journal} {J. Phys.: Condens. Matter.}\ }\textbf {\bibinfo
  {volume} {24}},\ \bibinfo {pages} {195104} (\bibinfo {year}
  {2012})}\BibitemShut {NoStop}%
\bibitem [{\citenamefont {Schir{\'o}}\ \emph {et~al.}(2011)\citenamefont
  {Schir{\'o}}, \citenamefont {Caronna}, \citenamefont {Natali}, \citenamefont
  {Koza},\ and\ \citenamefont {Cupane}}]{schiro11}%
  \BibitemOpen
  \bibfield  {author} {\bibinfo {author} {\bibfnamefont {G.}~\bibnamefont
  {Schir{\'o}}}, \bibinfo {author} {\bibfnamefont {C.}~\bibnamefont {Caronna}},
  \bibinfo {author} {\bibfnamefont {F.}~\bibnamefont {Natali}}, \bibinfo
  {author} {\bibfnamefont {M.~M.}\ \bibnamefont {Koza}}, \ and\ \bibinfo
  {author} {\bibfnamefont {A.}~\bibnamefont {Cupane}},\ }\href {\doibase
  10.1021/jz200797g} {\bibfield  {journal} {\bibinfo  {journal} {J. Phys. Chem.
  Lett.}\ }\textbf {\bibinfo {volume} {2}},\ \bibinfo {pages} {2275} (\bibinfo
  {year} {2011})}\BibitemShut {NoStop}%
\bibitem [{\citenamefont {Wang}\ \emph {et~al.}(1994)\citenamefont {Wang},
  \citenamefont {Sheng}, \citenamefont {Yang}, \citenamefont {Zhuang},
  \citenamefont {Lou},\ and\ \citenamefont {Chen}}]{wang94}%
  \BibitemOpen
  \bibfield  {author} {\bibinfo {author} {\bibfnamefont {W.~Q.}\ \bibnamefont
  {Wang}}, \bibinfo {author} {\bibfnamefont {X.~R.}\ \bibnamefont {Sheng}},
  \bibinfo {author} {\bibfnamefont {H.~S.}\ \bibnamefont {Yang}}, \bibinfo
  {author} {\bibfnamefont {Z.~H.}\ \bibnamefont {Zhuang}}, \bibinfo {author}
  {\bibfnamefont {F.~M.}\ \bibnamefont {Lou}}, \ and\ \bibinfo {author}
  {\bibfnamefont {Z.~J.}\ \bibnamefont {Chen}},\ }\href@noop {} {\bibfield
  {journal} {\bibinfo  {journal} {J. Biol. Phys.}\ }\textbf {\bibinfo {volume}
  {20}},\ \bibinfo {pages} {247} (\bibinfo {year} {1994})}\BibitemShut
  {NoStop}%
\bibitem [{\citenamefont {Lima}\ \emph {et~al.}(2001)\citenamefont {Lima},
  \citenamefont {Freire}, \citenamefont {Sasaki}, \citenamefont {Melo},
  \citenamefont {Mendes},\ and\ \citenamefont {Moreira}}]{lima01}%
  \BibitemOpen
  \bibfield  {author} {\bibinfo {author} {\bibfnamefont {R.~J.~C.}\
  \bibnamefont {Lima}}, \bibinfo {author} {\bibfnamefont {P.~T.~C.}\
  \bibnamefont {Freire}}, \bibinfo {author} {\bibfnamefont {J.~M.}\
  \bibnamefont {Sasaki}}, \bibinfo {author} {\bibfnamefont {F.~E.~A.}\
  \bibnamefont {Melo}}, \bibinfo {author} {\bibfnamefont {J.}~\bibnamefont
  {Mendes}}, \ and\ \bibinfo {author} {\bibfnamefont {R.~L.}\ \bibnamefont
  {Moreira}},\ }\href {\doibase 10.1002/jrs.739} {\bibfield  {journal}
  {\bibinfo  {journal} {J. Raman Spectrosc.}\ }\textbf {\bibinfo {volume}
  {32}},\ \bibinfo {pages} {751} (\bibinfo {year} {2001})}\BibitemShut
  {NoStop}%
\bibitem [{\citenamefont {Drebushchak}\ \emph {et~al.}(2005)\citenamefont
  {Drebushchak}, \citenamefont {Boldyreva}, \citenamefont {Kovalevskaya},
  \citenamefont {Paukov},\ and\ \citenamefont {Drebushchak}}]{drebushchak05}%
  \BibitemOpen
  \bibfield  {author} {\bibinfo {author} {\bibfnamefont {V.~A.}\ \bibnamefont
  {Drebushchak}}, \bibinfo {author} {\bibfnamefont {E.~V.}\ \bibnamefont
  {Boldyreva}}, \bibinfo {author} {\bibfnamefont {Y.~A.}\ \bibnamefont
  {Kovalevskaya}}, \bibinfo {author} {\bibfnamefont {I.~E.}\ \bibnamefont
  {Paukov}}, \ and\ \bibinfo {author} {\bibfnamefont {T.~N.}\ \bibnamefont
  {Drebushchak}},\ }\href {\doibase 10.1007/s10973-004-0563-8} {\bibfield
  {journal} {\bibinfo  {journal} {J. Therm. Anal. Calorim.}\ }\textbf {\bibinfo
  {volume} {79}},\ \bibinfo {pages} {65} (\bibinfo {year} {2005})}\BibitemShut
  {NoStop}%
\bibitem [{\citenamefont {Friedman}(1973)}]{friedman1973chemistry}%
  \BibitemOpen
  \bibfield  {author} {\bibinfo {author} {\bibfnamefont {M.}~\bibnamefont
  {Friedman}},\ }\href@noop {} {\emph {\bibinfo {title} {The chemistry and
  biochemistry of the sulfhydryl group in amino acids, peptides and
  proteins}}}\ (\bibinfo  {publisher} {Pergamon press Oxford},\ \bibinfo {year}
  {1973})\BibitemShut {NoStop}%
\bibitem [{\citenamefont {Paukov}\ \emph {et~al.}(2007)\citenamefont {Paukov},
  \citenamefont {Kovalevskaya}, \citenamefont {Drebushchak}, \citenamefont
  {Drebushchak},\ and\ \citenamefont {Boldyreva}}]{paukov07}%
  \BibitemOpen
  \bibfield  {author} {\bibinfo {author} {\bibfnamefont {I.~E.}\ \bibnamefont
  {Paukov}}, \bibinfo {author} {\bibfnamefont {Y.~A.}\ \bibnamefont
  {Kovalevskaya}}, \bibinfo {author} {\bibfnamefont {V.~A.}\ \bibnamefont
  {Drebushchak}}, \bibinfo {author} {\bibfnamefont {T.~N.}\ \bibnamefont
  {Drebushchak}}, \ and\ \bibinfo {author} {\bibfnamefont {E.~V.}\ \bibnamefont
  {Boldyreva}},\ }\href {\doibase 10.1021/jp073655w} {\bibfield  {journal}
  {\bibinfo  {journal} {J. Phys. Chem. B}\ }\textbf {\bibinfo {volume} {111}},\
  \bibinfo {pages} {9186} (\bibinfo {year} {2007})}\BibitemShut {NoStop}%
\bibitem [{\citenamefont {Paukov}\ \emph {et~al.}(2008)\citenamefont {Paukov},
  \citenamefont {Kovalevskaya},\ and\ \citenamefont {Boldyreva}}]{paukov08}%
  \BibitemOpen
  \bibfield  {author} {\bibinfo {author} {\bibfnamefont {I.~E.}\ \bibnamefont
  {Paukov}}, \bibinfo {author} {\bibfnamefont {Y.~A.}\ \bibnamefont
  {Kovalevskaya}}, \ and\ \bibinfo {author} {\bibfnamefont {E.~V.}\
  \bibnamefont {Boldyreva}},\ }\href {\doibase 10.1007/s10973-007-8697-0}
  {\bibfield  {journal} {\bibinfo  {journal} {J. Therm. Anal. Calorim.}\
  }\textbf {\bibinfo {volume} {93}},\ \bibinfo {pages} {423} (\bibinfo {year}
  {2008})}\BibitemShut {NoStop}%
\bibitem [{\citenamefont {Moggach}\ \emph {et~al.}(2005)\citenamefont
  {Moggach}, \citenamefont {Clark},\ and\ \citenamefont {Parsons}}]{moggach05}%
  \BibitemOpen
  \bibfield  {author} {\bibinfo {author} {\bibfnamefont {S.~A.}\ \bibnamefont
  {Moggach}}, \bibinfo {author} {\bibfnamefont {S.~J.}\ \bibnamefont {Clark}},
  \ and\ \bibinfo {author} {\bibfnamefont {S.}~\bibnamefont {Parsons}},\
  }\href@noop {} {\bibfield  {journal} {\bibinfo  {journal} {Acta Crystallogr.
  Sect. B-Struct. Sci.}\ }\textbf {\bibinfo {volume} {61}},\ \bibinfo {pages}
  {o2739} (\bibinfo {year} {2005})}\BibitemShut {NoStop}%
\bibitem [{\citenamefont {Roux}\ \emph {et~al.}(2010)\citenamefont {Roux},
  \citenamefont {Foces-Foces}, \citenamefont {Notario}, \citenamefont
  {Ribeiro~da Silva}, \citenamefont {Ribeiro~da Silva}, \citenamefont
  {Santos},\ and\ \citenamefont {Juaristi}}]{roux10}%
  \BibitemOpen
  \bibfield  {author} {\bibinfo {author} {\bibfnamefont {M.~V.}\ \bibnamefont
  {Roux}}, \bibinfo {author} {\bibfnamefont {C.}~\bibnamefont {Foces-Foces}},
  \bibinfo {author} {\bibfnamefont {R.}~\bibnamefont {Notario}}, \bibinfo
  {author} {\bibfnamefont {M.~A.~V.}\ \bibnamefont {Ribeiro~da Silva}},
  \bibinfo {author} {\bibfnamefont {M.~D. M.~C.}\ \bibnamefont {Ribeiro~da
  Silva}}, \bibinfo {author} {\bibfnamefont {A.~F. L. O.~M.}\ \bibnamefont
  {Santos}}, \ and\ \bibinfo {author} {\bibfnamefont {E.}~\bibnamefont
  {Juaristi}},\ }\href {\doibase 10.1021/jp1025637} {\bibfield  {journal}
  {\bibinfo  {journal} {The J. Phys. Chem. B}\ }\textbf {\bibinfo {volume}
  {114}},\ \bibinfo {pages} {10530} (\bibinfo {year} {2010})}\BibitemShut
  {NoStop}%
\bibitem [{\citenamefont {Larson}\ and\ \citenamefont
  {Von~Dreele}(1994)}]{larson94}%
  \BibitemOpen
  \bibfield  {author} {\bibinfo {author} {\bibfnamefont {A.~C.}\ \bibnamefont
  {Larson}}\ and\ \bibinfo {author} {\bibfnamefont {R.~B.}\ \bibnamefont
  {Von~Dreele}},\ }\href@noop {} {\bibfield  {journal} {\bibinfo  {journal}
  {Los Alamos National Laboratory Report LAUR 86-748}\ } (\bibinfo {year}
  {1994})}\BibitemShut {NoStop}%
\bibitem [{\citenamefont {Toby}(2001)}]{toby01}%
  \BibitemOpen
  \bibfield  {author} {\bibinfo {author} {\bibfnamefont {B.~H.}\ \bibnamefont
  {Toby}},\ }\href {\doibase 10.1107/S0021889801002242} {\bibfield  {journal}
  {\bibinfo  {journal} {J. Appl. Crystallogr.}\ }\textbf {\bibinfo {volume}
  {34}},\ \bibinfo {pages} {210} (\bibinfo {year} {2001})}\BibitemShut
  {NoStop}%
\bibitem [{\citenamefont {Kerr}\ \emph {et~al.}(1975)\citenamefont {Kerr},
  \citenamefont {Ashmore},\ and\ \citenamefont {Koetzle}}]{kerr75}%
  \BibitemOpen
  \bibfield  {author} {\bibinfo {author} {\bibfnamefont {K.~A.}\ \bibnamefont
  {Kerr}}, \bibinfo {author} {\bibfnamefont {J.~P.}\ \bibnamefont {Ashmore}}, \
  and\ \bibinfo {author} {\bibfnamefont {T.~F.}\ \bibnamefont {Koetzle}},\
  }\href {\doibase 10.1107/S0567740875006772} {\bibfield  {journal} {\bibinfo
  {journal} {Acta Crystallogr. B}\ }\textbf {\bibinfo {volume} {31}},\ \bibinfo
  {pages} {2022} (\bibinfo {year} {1975})}\BibitemShut {NoStop}%
\bibitem [{\citenamefont {Oughton}\ and\ \citenamefont
  {Harrison}(1959)}]{oughton59}%
  \BibitemOpen
  \bibfield  {author} {\bibinfo {author} {\bibfnamefont {B.~M.}\ \bibnamefont
  {Oughton}}\ and\ \bibinfo {author} {\bibfnamefont {P.~M.}\ \bibnamefont
  {Harrison}},\ }\href {\doibase 10.1107/S0365110X59001177} {\bibfield
  {journal} {\bibinfo  {journal} {Acta Crystallogr.}\ }\textbf {\bibinfo
  {volume} {12}},\ \bibinfo {pages} {396} (\bibinfo {year} {1959})}\BibitemShut
  {NoStop}%
\bibitem [{\citenamefont {Lashley}\ \emph {et~al.}(2003)\citenamefont
  {Lashley}, \citenamefont {Hundley}, \citenamefont {Migliori}, \citenamefont
  {Sarrao}, \citenamefont {Pagliuso}, \citenamefont {Darling}, \citenamefont
  {Jaime}, \citenamefont {Cooley}, \citenamefont {Hults}, \citenamefont
  {Morales}, \citenamefont {Thoma}, \citenamefont {Smith}, \citenamefont
  {Boerio-Goates}, \citenamefont {Woodfield}, \citenamefont {Stewart},
  \citenamefont {Fisher},\ and\ \citenamefont {Phillips}}]{lashley03}%
  \BibitemOpen
  \bibfield  {author} {\bibinfo {author} {\bibfnamefont {J.~C.}\ \bibnamefont
  {Lashley}}, \bibinfo {author} {\bibfnamefont {M.~F.}\ \bibnamefont
  {Hundley}}, \bibinfo {author} {\bibfnamefont {A.}~\bibnamefont {Migliori}},
  \bibinfo {author} {\bibfnamefont {J.~L.}\ \bibnamefont {Sarrao}}, \bibinfo
  {author} {\bibfnamefont {P.~G.}\ \bibnamefont {Pagliuso}}, \bibinfo {author}
  {\bibfnamefont {T.~W.}\ \bibnamefont {Darling}}, \bibinfo {author}
  {\bibfnamefont {M.}~\bibnamefont {Jaime}}, \bibinfo {author} {\bibfnamefont
  {J.~C.}\ \bibnamefont {Cooley}}, \bibinfo {author} {\bibfnamefont {W.~L.}\
  \bibnamefont {Hults}}, \bibinfo {author} {\bibfnamefont {L.}~\bibnamefont
  {Morales}}, \bibinfo {author} {\bibfnamefont {D.~J.}\ \bibnamefont {Thoma}},
  \bibinfo {author} {\bibfnamefont {J.~L.}\ \bibnamefont {Smith}}, \bibinfo
  {author} {\bibfnamefont {J.}~\bibnamefont {Boerio-Goates}}, \bibinfo {author}
  {\bibfnamefont {B.~F.}\ \bibnamefont {Woodfield}}, \bibinfo {author}
  {\bibfnamefont {G.~R.}\ \bibnamefont {Stewart}}, \bibinfo {author}
  {\bibfnamefont {R.~A.}\ \bibnamefont {Fisher}}, \ and\ \bibinfo {author}
  {\bibfnamefont {N.~E.}\ \bibnamefont {Phillips}},\ }\href {\doibase
  10.1016/S0011-2275(03)00092-4} {\bibfield  {journal} {\bibinfo  {journal}
  {Cryogenics}\ }\textbf {\bibinfo {volume} {43}},\ \bibinfo {pages} {369}
  (\bibinfo {year} {2003})}\BibitemShut {NoStop}%
\bibitem [{\citenamefont {Elliott}(1992)}]{elliott1992unified}%
  \BibitemOpen
  \bibfield  {author} {\bibinfo {author} {\bibfnamefont {S.}~\bibnamefont
  {Elliott}},\ }\href@noop {} {\bibfield  {journal} {\bibinfo  {journal} {EPL
  (Europhys. Lett.)}\ }\textbf {\bibinfo {volume} {19}},\ \bibinfo {pages}
  {201} (\bibinfo {year} {1992})}\BibitemShut {NoStop}%
\bibitem [{\citenamefont {Caride}\ and\ \citenamefont
  {Tsallis}(1984)}]{caride84}%
  \BibitemOpen
  \bibfield  {author} {\bibinfo {author} {\bibfnamefont {A.~O.}\ \bibnamefont
  {Caride}}\ and\ \bibinfo {author} {\bibfnamefont {C.}~\bibnamefont
  {Tsallis}},\ }\href {\doibase 10.1007/BF01017374} {\bibfield  {journal}
  {\bibinfo  {journal} {J. Stat. Phys.}\ }\textbf {\bibinfo {volume} {35}},\
  \bibinfo {pages} {187} (\bibinfo {year} {1984})}\BibitemShut {NoStop}%
\bibitem [{\citenamefont {Onsager}(1944)}]{onsager44}%
  \BibitemOpen
  \bibfield  {author} {\bibinfo {author} {\bibfnamefont {L.}~\bibnamefont
  {Onsager}},\ }\href {\doibase 10.1103/PhysRev.65.117} {\bibfield  {journal}
  {\bibinfo  {journal} {Phys. Rev.}\ }\textbf {\bibinfo {volume} {65}},\
  \bibinfo {pages} {117} (\bibinfo {year} {1944})}\BibitemShut {NoStop}%
\bibitem [{\citenamefont {Shuker}\ and\ \citenamefont
  {Gammon}(1970)}]{shuker1970raman}%
  \BibitemOpen
  \bibfield  {author} {\bibinfo {author} {\bibfnamefont {R.}~\bibnamefont
  {Shuker}}\ and\ \bibinfo {author} {\bibfnamefont {R.~W.}\ \bibnamefont
  {Gammon}},\ }\href@noop {} {\bibfield  {journal} {\bibinfo  {journal} {Phys.
  Rev. Lett.}\ }\textbf {\bibinfo {volume} {25}},\ \bibinfo {pages} {222}
  (\bibinfo {year} {1970})}\BibitemShut {NoStop}%
\bibitem [{\citenamefont {Ahmad}\ \emph {et~al.}(1986)\citenamefont {Ahmad},
  \citenamefont {Hutt},\ and\ \citenamefont {A}}]{ahmad1986}%
  \BibitemOpen
  \bibfield  {author} {\bibinfo {author} {\bibfnamefont {N.}~\bibnamefont
  {Ahmad}}, \bibinfo {author} {\bibfnamefont {K.~W.}\ \bibnamefont {Hutt}}, \
  and\ \bibinfo {author} {\bibfnamefont {P.~W.}\ \bibnamefont {A}},\
  }\href@noop {} {\bibfield  {journal} {\bibinfo  {journal} {J. Phys. C: Solid
  State Phys.}\ }\textbf {\bibinfo {volume} {19}},\ \bibinfo {pages} {3765}
  (\bibinfo {year} {1986})}\BibitemShut {NoStop}%
\bibitem [{\citenamefont {Binder}\ and\ \citenamefont {Kob}(2006)}]{binder}%
  \BibitemOpen
  \bibfield  {author} {\bibinfo {author} {\bibfnamefont {K.}~\bibnamefont
  {Binder}}\ and\ \bibinfo {author} {\bibfnamefont {W.}~\bibnamefont {Kob}},\
  }\href@noop {} {\emph {\bibinfo {title} {Glassy materials and disordered
  solids, an introduction to their statistical mechanics}}}\ (\bibinfo
  {publisher} {World Scientific Publishing Co. Pte. Ltd.},\ \bibinfo {year}
  {2006})\ pp.\ \bibinfo {pages} {203--210}\BibitemShut {NoStop}%
\end{thebibliography}%
\end{document}